\documentclass[11pt]{pazh}

\usepackage[T2A]{fontenc}
\usepackage[utf8]{inputenc}

\usepackage{latexsym}
\usepackage{amssymb}

\usepackage{graphics}
\usepackage{graphicx}
\usepackage{latexsym}
\usepackage{amssymb}
\usepackage{amsmath}
\usepackage{hyperref}
\usepackage[hyphenbreaks]{breakurl}

\begin{document}

\sloppypar

\title{\bf Spatial environment of polar-ring galaxies from the SDSS}

\author{S.S. Savchenko\inst{1}, V.P. Reshetnikov\inst{1}} 

\institute{St.Petersburg State University, Universitetskii pr. 28, St.Petersburg, 
198504 Russia\\
\hspace*{3cm}savchenko.s.s@gmail.com, v.reshetnikov@spbu.ru
}

\titlerunning{Environment of PRGs}

\abstract{Based on SDSS data, we have considered the spatial environment of 
galaxies with extended polar rings. We used two approaches: estimating the 
projected distance to the nearest companion and counting the number of companions 
as a function of the distance to the galaxy. Both approaches have
shown that the spatial environment of polar-ring galaxies on scales of hundreds 
of kiloparsecs is, on average, less dense than that of galaxies without polar 
structures. Apparently, one of the main causes of this effect is that the polar 
structures in a denser environment are destroyed more often during encounters
and mergers with other galaxies.
}
\titlerunning{Environment of PRGs}
\maketitle

\section{Introduction}

Polar-ring galaxies (PRGs) are very rare and interesting extragalactic objects. 
Two large-scale subsystems coexist in their structure: a central galaxy
and a ring or disk oriented at a large angle to its major axis (the catalog by 
Whitmore et al. 1990; the SPRC catalog by Moiseev et al. 2011). As a
rule, the host galaxies and the polar structures differ noticeably in their 
characteristics. The central galaxies in most PRGs are gas-poor early-type (E/S0)
galaxies. By contrast, the polar structures are usually gas-rich, have blue colors, 
exhibit star formation, and generally resemble the disks of spiral galaxies (see,
e.g., Combes 2006; Reshetnikov \& Combes 2015; Moiseev et al. 2015; and references 
therein). 

Some ``secondary'' event in the PRG history is usually invoked to 
explain the formation of kinematically and morphologically decoupled structures. 
Various secondary events are considered: the capture of matter from an approached 
galaxy (Reshetnikov \& Sotnikova 1997; Bournaud \& Combes 2003), the merging of 
galaxies (Bekki 1998; Bournaud \& Combes 2003), and the accretion of matter 
from intergalactic space (Maccio et al. 2006; Brook et al. 2008). Observations 
show that several PRG formation mechanisms can apparently be realized, but the
relative contribution of different mechanisms remains unclear.

One test of the formation models of PRGs is a statistical study of their spatial 
environment. (For example, if the polar rings are formed mainly during
close encounters of galaxies, then one might expect the PRGs to have an excess 
of close neighbors.) Unfortunately, such studies are very rare so far. Brocca
et al. (1997) considered objects from the catalog by Whitmore et al. (1990) and 
found the spatial environment of PRGs to be similar to that of normal galaxies.
From this analysis the authors concluded that if the polar structures are formed 
during the interaction of galaxies, then these processes mainly occurred very
long (billions of years) ago and by now the signatures of interactions and mergers 
in the PRG environment have been washed out. Finkelman et al. (2012)
studied the membership of PRGs selected from the SPRC in known groups of galaxies 
and concluded that, on average, the PRGs are located in a less dense
spatial environment than are the ordinary early-type galaxies.

The goal of this note is to study the spatial environment of various groups of 
galaxies from the SPRC. The SPRC catalog was compiled on the basis of the 
Sloan Digital Sky Survey (SDSS), and it is much more homogeneous than the catalog by
Whitmore et al. (1990), which allows the nearest neighborhood of PRGs to be 
investigated quite comprehensively.

All numerical values in the paper are given for the cosmological model with the Hubble
constant of 70 km s$^{-1}$ Mpc$^{-1}$ and $\Omega_m=0.3$, $\Omega_{\Lambda}=0.7$.

\section{Analysis of the spatial environment}

\subsection{Samples of galaxies}

The SPRC catalog (Moiseev et al. 2011) contains 275 galaxies divided into four groups. 
The first group includes 70 objects that are the best PRG candidates. They are 
morphologically similar to the classical PRGs from the catalog by Whitmore et al.
(1990). The preceding experience of studying such galaxies has shown that almost 
all objects with such a morphology are PRGs and contain two large-scale
kinematically decoupled subsystems (for examples see fig.~1 in Moiseev et al. (2015)). 
Below this group will be referred to as the B (best) sample.

The second group of SPRC objects includes 115 galaxies that are good PRG candidates. 
We will designate this sample as G (good).

The third and fourth parts of the SPRC catalog contain galaxies that may be related 
to PRGs (the R (related) sample, 53 galaxies) and galaxies where the presumed polar 
ring is seen nearly face-on (the P (possible face-on rings) sample, 37 galaxies).

A detailed description of all groups of galaxies is given in the SPRC.

\subsection{Methods for studying the environment of galaxies}

We applied two different approaches to study the environment of PRGs from the SDSS: 
estimating the distance to the nearest galaxy (companion) and counting the number 
of companions as a function of the distance to the galaxy.

We used the following criteria to classify a galaxy as a companion.

(1) The apparent $r$ magnitude of the companion should not be fainter than the 
apparent magnitude of the main galaxy increased by 2$^m$: $m_{comp} \leq m_{main} + 2^m$.

(2) The redshift difference between the galaxy being studied and its companion 
should not exceed 0.001 (300 km/s) or 0.0015 (450 km/s).

(3) The distance between the galactic centers in projection onto the plane of the 
sky should be less than a preset value dependent on the approach used (see below).

We searched for companion galaxies using the SQL queries to the SDSS
server\footnote{\burl{https://skyserver.sdss.org/dr12/en/tools/search/x_results.aspx }} 
in which the above constraints were specified.

The $r$-band data on galaxies were taken from the SDSS DR 12 (Alam et al. 2015). 
To avoid the errors due to the SDSS incompleteness, we took
only objects brighter than $r = 16^m$ from the SPRC. Including less bright galaxies 
in the sample does not allow us to search for faint companions by the methods
described below because of the SDSS limitation on the completeness of spectroscopic 
data for galaxies (Strauss et al. 2002). As a result, 22, 20, 29,
and 12 galaxies were included in the B, G, R, and P samples, respectively. \\

{\bf I. Distance to the nearest neighbor.}

In this approach the distance between the galaxy being studied and its nearest 
companion (the nearest galaxy satisfying the above magnitude and redshift criteria) 
was used as an estimate of the spatial density of galaxies. As the distance we used 
the distance projected onto the plane of the sky expressed either in Petrosian radii
($R_{petro}$) of the central galaxy or in kiloparsecs. In the former and latter 
cases, we searched for companions within 100$R_{petro}$ and 500 kpc, respectively. 
If no neighbors were found within these regions, then for such galaxies we took 
100$R_{petro}$ or 500 kpc, respectively, as an estimate of the distance to the companion
(i.e., we used the lower limit). \\

{\bf II. Dependence of the number of companions on distance.}

In this test for each galaxy from the sample we determined the number of companions as
a function of the distance to the central galaxy. To this end, for each galaxy we 
found the number of companions within 2, 4, 8, 16, 32, and 64 Petrosian
radii and, as a separate test, within 10, 20, 40, 80, 160, and 320 kpc. \\

{\bf III. Comparison sample.}

To compare the results of our study of the spatial environment for PRGs
and ordinary galaxies, we produced a comparison sample. The following principle 
was used to produce the comparison sample. For each galaxy of the
SPRC catalog we selected galaxies from the entire SDSS with close apparent radii, 
apparent magnitudes, colors, and redshifts: 
$\Delta R_{petro} < 15\%$, $\Delta m < 0.15^m$, $\Delta (g-r) < 0.05^m$,
$\Delta z < 10 \%$.
The limitation in color is important for selecting galaxies of
similar types, because the spatial environment depends on the galaxy type 
(see, e.g., Skibba et al. 2009). We added the limitation in redshift in order
that the luminosity distributions of SPRC galaxies and comparison galaxies be also 
similar. On average, about 200 similar galaxies were found for each SPRC
galaxy (although this number could be considerably smaller for some galaxies). 
Thus, we obtained an expanded comparison sample that is larger in volume
than the SPRC sample by hundreds of times.

We then drew the working comparison samples from the expanded comparison sample 
by randomly selecting only one similar galaxy from the entire group
for each SPRC galaxy. As a result, not only the volume of the comparison sample 
is equal to the volume the SPRC catalog, but also the magnitude, redshift,
size, luminosity, and color distributions turn out to be close. Multiple repetition 
of this step allows us to obtain a large number of comparison samples, which
makes it possible to perform a series of tests on them, to average the results, 
and to estimate the errors.

\section{Results and discussion}

\subsection{Minimum distance to the nearest neighbor}

Tables 1 and 2 present the results of our test to determine the distance to the 
nearest galaxy expressed, respectively, in Petrosian radii (Table 1) and kiloparsecs
(Table 2) for two limitations in redshift. The second and fourth columns of the 
tables give the mean values of this parameter for each galaxy subtype from
the SPRC (B, G, R, and P) and for the comparison sample (the cmp row). In addition, 
we combined the B and G types of the most reliable candidates into the
BG type to increase the number of galaxies in this group, thereby increasing the 
statistical weight of the result. The third and fifth columns of Tables 1 and 2
give the root-mean-square (rms) deviations $\sigma$; for the subgroups of the SPRC catalog 
this is the rms deviation from the mean for galaxies in the group, while for
the comparison sample this is the rms deviation from the mean for 50 realizations. 
The nearly zero scatter in the control group is the result of averaging over
many realizations of working comparison samples: the term $N^{-\frac{1}{2}}$ 
in the formula for the rms deviation led to a small resulting scatter.

\begin{table}[h]
\caption{Mean distance to the nearest companion expressed in Petrosian radii of 
the central galaxy for $\Delta\,z \leq 0.001$ (column 2) and 
$\Delta\,z \leq 0.0015$ (column 4)} 
\begin{center}
\begin{tabular}{| l | c c  | c c |}
\hline
Type & $\left< d\right>[R_{petro}]$ & $\sigma$ & $\left< d\right>[R_{petro}]$  & $\sigma$  \\ 
\hline
B   &  74.5  &  1.6 & 71.9 & 1.5\\
G   &  74.1  &  1.7 & 73.9 & 1.7\\
BG  &  74.3  &  0.8 & 72.8 & 0.8\\
R   &  61.3  &  1.2 & 57.5 & 1.2\\
P   &  57.9  &  3.2 & 57.9 & 3.2\\
cmp &  45.6  &  0.0 & 43.4 & 0.0\\
\hline
\end{tabular}
\end{center}
\label{tabl1}
\end{table}

\begin{table}[h]
\caption{Same as Table~1 expressed in kiloparsecs} 
\begin{center}
\begin{tabular}{| l |  c c  | c c | }
\hline
Type & $\left< d\right>[kpc]$  & $\sigma$ & $\left< d\right>[kpc]$  & $\sigma$ \\ 
\hline
B   &  402.7  &  5.9 &  402.7  & 5.9\\
G   &  371.3  &  8.1 &  370.7  & 8.1\\
BG  &  387.7  &  3.5 &  387.5  & 3.5 \\
R   &  393.4  &  5.7 &  388.3  & 5.6\\
P   &  297.2  &  15.5&  297.2  & 15.5 \\
cmp &  346.8  &  0.0 &  335.1  & 0.0 \\
\hline
\end{tabular}
\end{center}
\label{tabl2}
\end{table}

It can be seen from Tables 1 and 2 that the mean distance to the nearest galaxy 
for the best PRG candidates is larger than that for the less reliable candidates
and for the galaxies of the comparison sample, with this dependence being more 
pronounced when measuring the distances in Petrosian radii, i.e., in
the relative scale that takes into account the galaxy sizes. As expected, using 
the weaker limitation on the redshift difference ($\Delta\,z \leq 0.0015$) 
reduces the distance to the nearest galaxy but does not change the
above trend. Thus, according to this test, the polar-ring galaxies are, on average, 
in a less dense spatial environment than are similar galaxies without polar
structures.

\subsection{Dependence of the number of companions on distance}

The dependence of the mean number of companions on the distance to the galaxy 
for various subgroups of the SPRC catalog and for the comparison
sample is presented in Fig.~1 (for the distances expressed in Petrosian radii) 
and Fig.~2 (in kiloparsecs). The top and bottom panels of these figures show the
results obtained for the limitations on the redshift difference 
$\Delta\,z \leq 0.001$ and $\Delta\,z \leq 0.0015$, respectively.

\begin{figure*}
\centering
\includegraphics[width=0.8\textwidth, angle=0, clip=]{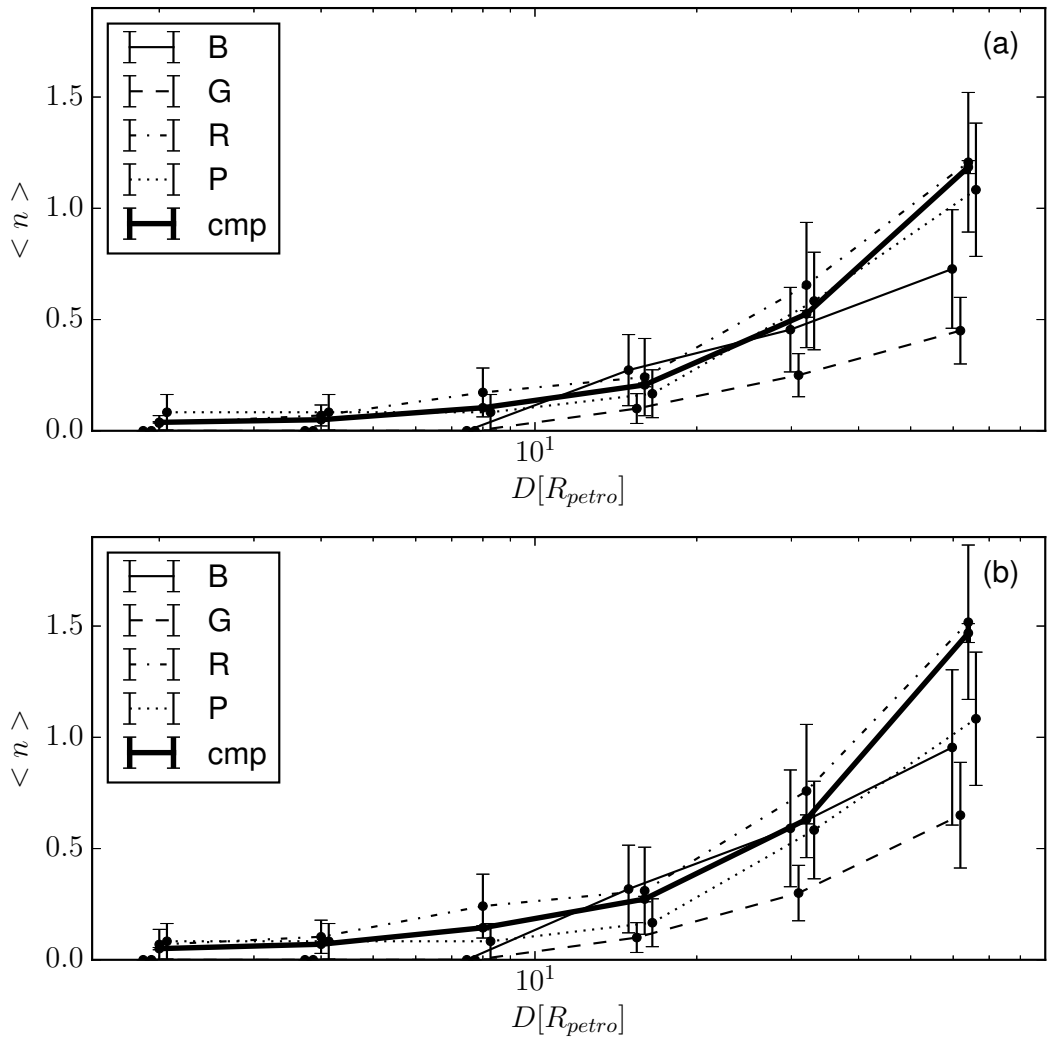}\\
\caption{Mean number of companions versus distance to the galaxy for various 
subgroups of the SPRC catalog and for the comparison sample. The results of our 
test for the $\Delta\,z$ limits of (a) 0.001 and (b) 0.0015. The distance to the 
galaxy is expressed in Petrosian radii. The data points are slightly displaced 
along the horizontal axis to avoid an overlap between the
error bars indicating the rms deviation.}
\label{fig1}
\end{figure*}

\begin{figure*}
\centering
\includegraphics[width=0.8\textwidth, angle=0, clip=]{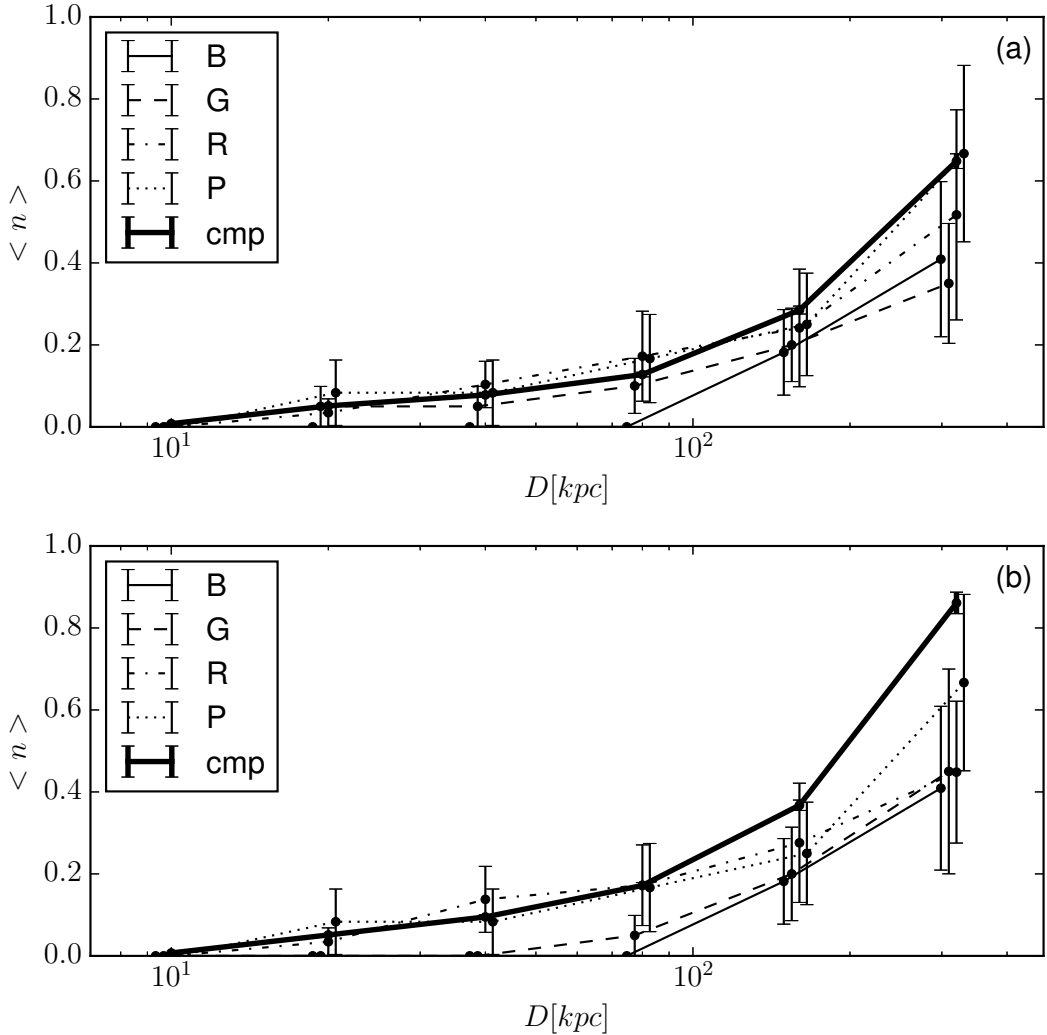}\\
\caption{Same as Fig.~1, but the distances are expressed in kiloparsecs.}
\label{fig2}
\end{figure*}

It can be seen from the figures that the dependences for all galaxy types do 
not differ statistically at relatively small distances (tens of $R_{petro}$, 
tens of kiloparsecs). This may be due to the small volume of the PRG samples and, 
accordingly, the small number of PRG companions.

As the distance increases, the dependences for the most probable PRG candidates 
(the B and G subgroups) begin to deviate from the dependence for
the comparison sample toward a smaller number of companions. This effect is most 
prominent in Fig.~2, where the distances are expressed in kiloparsecs.
Both figures also clearly show a trend where the dependences for the objects of 
the B and G subgroups at large distances lie below the dependences for other
subgroups and for the comparison sample.

As with the previous test, the results of our analysis suggest that the spatial 
environment of PRGs turns out to be less dense. The mean number of
companions expectedly increases for the tests with a weaker redshift constraints, 
but the overall character of the dependences is retained or even enhanced.

\section{Conclusions}

Using two simple and clear approaches, we investigated the environment of polar-ring 
galaxies from the SDSS and obtained the following main results.

(1) The mean projected distance to the nearest companion galaxy for PRG candidates 
is larger than that for galaxies without polar structures. For example,
this difference for the B subgroup reaches a factor of 1.6 for the distance 
expressed in Petrosian radii.

(2) The mean number of companion galaxies within several hundred kiloparsecs for 
PRG candidates is smaller than that for ordinary galaxies.
These results are consistent with the conclusion reached by Finkelman et al. (2012), 
who studied the occurrence of PRGs in galaxy groups and found that
the PRGs are predominantly located in a less dense environment than are the normal 
galaxies. On the other hand, Brocca et al. (1997) previously found no
differences in the environment of PRGs and galaxies without polar structures. 
This may be because the catalog by Whitmore et al. (1990) used by them,
which includes a large number of objects that are not PRGs, is less homogeneous.

Of course, the less dense environment of PRGs needs to be confirmed further based 
on a much larger (and so far lacking) observational material. Nevertheless,
the agreement between the results obtained by different methods in our paper and 
in Finkelman et al. (2012) allows this conclusion to be considered
quite plausible.

How can the observed spatial environment of PRGs be explained? On the one hand, 
the polar structures can undoubtedly be formed during close encounters of galaxies. 
Several such objects, in which the accretion of matter from one galaxy onto
another and the formation of a circumpolar structure are observed directly, are 
well known (see, e.g., Reshetnikov et al. 1996; Cox et al. 2001; Keel 2004).
However, such structures are apparently relatively short-lived ones, because they 
will be destroyed during the interaction with neighboring galaxies and
during the capture of companions. For example, the long-term evolution of the 
polar structure formed by external accretion from the intergalactic medium
was traced in the cosmological numerical simulations by Maccio et al. (2006). 
This structure was destroyed by the dynamical effect of the companion merging
with the main galaxy approximately one billion years after its formation. When 
the authors removed all companions from the environment of the galaxy under
consideration, the polar ring existed in their simulations much longer. It is 
this effect that possibly leads to the observed reduced spatial density of galaxies in
the PRG environment; the polar structures have more chances to ``survive'' in a less 
dense environment.

On the other hand, the polar rings can be formed through the so-called cool 
($T \sim 10^4$ K) accretion of gas from filaments in the intergalactic medium (Maccio
et al. 2006; Connors et al. 2006; Brook et al. 2008). This scenario and the 
formation of a massive polar ring require prolonged (billions of years) and
``coherent'' accretion of matter (Snaith et al. 2012); the latter is more probable 
in regions with a low density of galaxies. (The influence of nearby galaxies can
reorient or even destroy the external accretion flow.) In addition, as has been 
noted above, the formed polar structure will be able to exist longer in a less dense
environment.

\bigskip
\section*{Acknowledgments}
This study is based on publicly SDSS data. SDSS is managed by the Astrophysical Research
Consortium for the Participating Institutions of the SDSS Collaboration including 
the Brazilian Participation Group, the Carnegie Institution for Science,
Carnegie Mellon University, the Chilean Participation Group, the French Participation 
Group, Harvard-Smithsonian Center for Astrophysics, Instituto
de Astrofisica de Canarias, the Johns Hopkins University, Kavli Institute for the 
Physics and Mathematics of the Universe (IPMU)/University of Tokyo,
Lawrence Berkeley National Laboratory, Leibniz Institut fur Astrophysik Potsdam (AIP), 
Max-Planck-Institut fur Astronomie (MPIA Heidelberg), Max-Planck-Institut fur 
Astrophysik (MPA Garching), Max-Planck-Institut fur Extraterrestrische Physik
(MPE), National Astronomical Observatories of China, New Mexico State University, 
New York University, University of Notre Dame, Observatorio Nacional/MCTI, the Ohio 
State University, Pennsylvania State University, Shanghai Astronomical
Observatory, United Kingdom Participation Group, Universidad Nacional Autonoma de 
Mexico, University of Arizona, University of Colorado Boulder,
University of Oxford, University of Portsmouth, University of Utah, University of 
Virginia, University of Washington, University of Wisconsin, Vanderbilt
University, and Yale University.

\section*{REFERENCES}

\hspace{0.4cm} 1. S. Alam, F.D. Albareti, C.A. Prieto, F. Anders,
S.F. Anderson, T. Anderton, B.H. Andrews, E. Armengaud,
et al., Astrophys. J. Suppl. Ser. 219, 12A (2015).

2. K. Bekki, Astrophys. J. 499, 635 (1998).

3. F. Bournaud and F. Combes, Astron. Astrophys. 401,
817 (2003).

4. Ch. Brocca, D. Bettoni, and G. Galletta, Astron.
Astrophys. 326, 907 (1997).

5. Ch.B. Brook, F. Governato, Th. Quinn, J. Wadsley,
A.M. Brooks, B. Willman, A. Stilp, and P. Jonsson,
Astrophys. J. 689, 678 (2008).

6. F. Combes, EAS Publ. Ser. 20, 97 (2006).

7. T.W. Connors, D. Kawata, J. Bailin, J. Tumlinson,
and B.K. Gibson, Astrophys. J. 646, L53 (2006).

8. A.L. Cox, L.S. Sparke, A.M. Watson, and
G. van Moorsel, Astron. J. 121, 692 (2001).

9. I. Finkelman, J.G. Funes, and N. Brosch, Mon. Not.
R. Astron. Soc. 422, 2386 (2012).

10. W.C. Keel, Astron. J. 127, 1325 (2004).

11. A.V. Maccio, B. Moore, and J. Stadel, Astrophys. J.
636, L25 (2006).

12. A.V. Moiseev, K.I. Smirnova, A.A. Smirnova, and
V.P. Reshetnikov, Mon.Not. R. Astron. Soc. 418, 244
(2011) (SPRC).

13. A. Moiseev, S. Khoperskov, A. Khoperskov,
K. Smirnova, A. Smirnova, A. Saburova, and
V. Reshetnikov, Baltic Astron. 24, 76 (2015).

14. V.P. Reshetnikov, V.A. Hagen-Thorn, and
V.A. Yakovleva, Astron. Astrophys. 314, 729
(1996).

15. V. Reshetnikov and F. Combes, Mon. Not. R. Astron.
Soc. 447, 2287 (2015).

16. V. Reshetnikov and N. Sotnikova, Astron. Astrophys.
325, 933 (1997).

17. R.A. Skibba, S.P. Bamford, R.C. Nichol, C.J. Lintott,
D. Andreescu, E.M. Edmondson, P. Murray,
M.J. Raddick, et al., Mon. Not. R. Astron. Soc. 399,
966 (2009).

18. O.N. Snaith, B.K. Gibson, C.B. Brook, A. Knebe,
R.J. Thacker, T.R. Quinn, F. Governato, and
P.B. Tissera, Mon. Not. R. Astron. Soc. 425, 1967
(2012).

19. M.A. Strauss, D.H. Weinberg, and R.H. Lupton,
Astron. J. 124, 1810 (2002).

20. B.C. Whitmore, R.A. Lucas, D.B. McElroy,
T.Y. Steiman-Cameron, P.D. Sackett, and
R.P. Olling, Astron. J. 100, 1489 (1990).

\end{document}